\documentclass[12pt,a4paper,final]{iopart}

\usepackage{iopams} 
\expandafter\let\csname equation*\endcsname\relax
\expandafter\let\csname endequation*\endcsname\relaxvariance
\usepackage{amsmath}
\usepackage{graphicx}
\usepackage{cite}
\usepackage[breaklinks=true,colorlinks=true,linkcolor=blue,urlcolor=blue,citecolor=blue]{hyperref}
\begin{document}

%
\title[Qubit transformations on Rashba ring]{Qubit transformations on Rashba ring with periodic potential}

\author{Ambro\v{z} Kregar$^{1,2}$ and Anton Ram\v sak$^{3}$}
\address{$^1$ Faculty of Mechanical Engineering, University of Ljubljana, A\v{s}ker\v{c}eva cesta 6, 1000 Ljubljana, Slovenia}
\address{$^2$ Institute of Physical and Theoretical Chemistry, Graz University of Technology, Stremayrgasse 9, A-8010 Graz, Austria}
\address{$^3$ Faculty of Mathematics and Physics, University of Ljubljana, Jadranska ulica 19, 1000 Ljubljana, Slovenia}
\ead{ambroz.kregar@fs.uni-lj.si}


\vspace{10pt}
\begin{indented}
\item[]8 April 2020
\end{indented}


\begin{abstract}
A spin-qubit transformation protocol is proposed for an electron in a mesoscopic quantum ring with tunable Rashba interaction controlled by the external electric field. The dynamics of an electron driven around the ring by a series of Landau-Zenner-like transitions between a finite number of local voltage gates is determined analytically. General single-qubit transformations are demonstrated to be feasible in a dynamical basis of localized pseudo-spin states. It is also demonstrated that by the use of suitable protocols based on changes of the Rashba interaction full Bloch sphere can be covered. The challenges of a possible realization of the proposed system in semiconductor heterostructures are discussed.\end{abstract}
%

\section{Introduction}

The spintronics, a promising new branch of electronics based on electron's spin as the information carrier instead of its charge, has emerged in the last few decades. The use of spin promises several important advantages in information processing, most notably longer coherence times and lower power consumption compared to classical electronic devices \cite{Wolf2001,Zutic2004,Rashba2007}. What is even more important is that the spintronic devices are among the most promising candidates for the realization of quantum computers with spin states being used as qubits \cite{Awschalom2013}. To avoid the use of the magnetic field for spin manipulation, the spin-orbit interaction (SOI) \cite{Winkler2003,Engel2007} might be used to control electron's spin. 
Rashba type SOI \cite{Rashba1960}, emerging as a consequence of structural inversion asymmetry of the effective potential in the semiconductor heterostructure, seems especially promising for this task since its magnitude can be artificially controlled by applying the external electric field perpendicular to the plane of the heterostructure \cite{Nitta1997,Schapers1998}. Potential use of this phenomenon was first demonstrated by SOI field effect transistor, proposed by Datta in 1990 \cite{Datta1990}, followed by several other proposals for two-dimensional spintronic devices\cite{Nitta1999,Wolf2001,Schliemann2003,Zutic2004,Wunderlich2010,Stajic2013}.

For the use in quantum computation, the spin transformation would ideally be applied to a single-electron qubit, trapped in a quantum dot, with its position determined by an external electric potential \cite{Ramsak2018}. Spin transformation for an arbitrary motion of an electron in one dimension system can be expressed analytically \cite{Cadez2013,Cadez2014} which also allows for exact analysis of errors in qubit transformations due to the noise in driving fields \cite{Ulcakar2017} and the effects of finite temperature \cite{Donvil2020}. Note, however, that since the Rashba spin rotation axis in this system is perpendicular to the direction of electrons' motion, one-dimensional motion provides only a limited range of possible spin transformations \cite{Ramsak2018}.

\begin{figure}[ht]
\centerline{\includegraphics[scale=1]{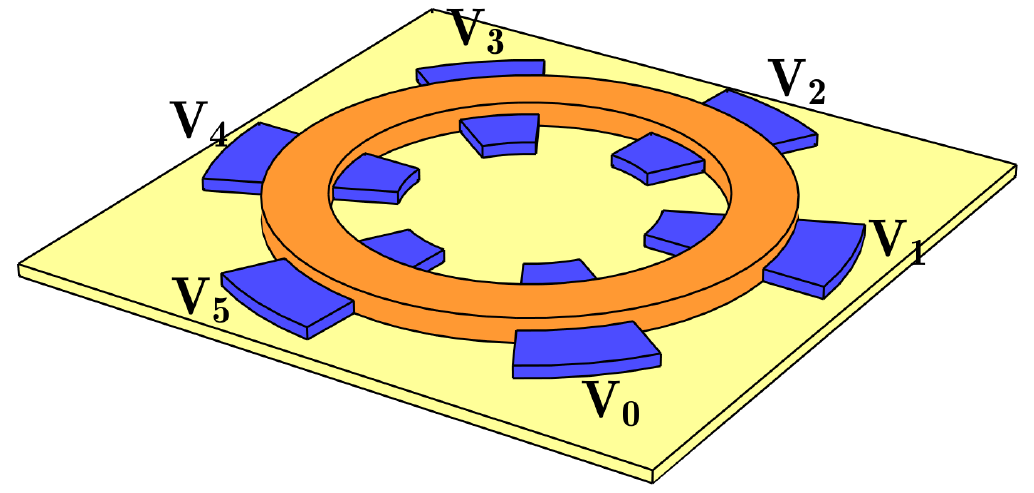}}
\caption[Ring-schematic]{Schematic representation of a quantum ring device with six voltage gates used to control the electron position.}
\label{fig:ring1}
\end{figure}

This limitation is removed by allowing the electron to move in two dimensions\cite{San-Jose2008,Golovach2010}. The system of electron on a quantum ring with the Rashba coupling is particularly convenient in this regard since it allows for the study of spin transformations in a two-dimensional system using effectively one-dimensional Hamiltonian \cite{Meijer2002}. As shown in Ref.~\cite{Kregar2016}, the motion of the electron around the ring with the Rashba coupling, tuned using external gate voltage, can be used to realize an arbitrary single-qubit transformation in the qubit basis of Kramers states. However, the authors assumed that the position of external potential can be shifted for an arbitrary azimuthal angle, which is usually not the case in realistic spintronic devices, where the potential is typically defined using fixed external voltage gates, applied to the surface of the semiconductor, as shown in figure \ref{fig:ring1}. The minima of the potential can, therefore, occur only at specific positions. To describe more realistic devices, this limiting factor should be taken into account.

The goal of this paper is to analyze the transformation of electron's spin state when transferred from the site of one voltage gate to the site of its neighboring gate. In the case of equidistant gates, forming a periodic potential, this can be done analytically. As we show in this paper, the spin rotation is directly related to the spin-dependent part of the hopping parameter, coupling the neighboring Wannier states in the corresponding tight-binding model of periodic gate potential. To find an explicit analytic form of hopping terms, we first calculate the Bloch functions on the ring, characterized by specific site-dependent Rashba-induced spin orientation, and their energies. Corresponding Wannier states and their nearest-neighbor hopping Hamiltonian, obtained by Fourier transformation of Bloch states and energies, are further transformed by local spin rotations to obtain a basis of localized states, resembling the pure spin state of the electron, trapped at the site of each voltage gate. The hopping terms between the states of this so-called spin Wannier basis is then expressed analytically by spin-rotation matrices, allowing a simple analysis of spin transformations accompanying electron transition.

The results are verified by numerical calculation of spin rotation during the slow transition of the electron between gates, showing that the use of Wannier hopping terms indeed results in correct spin transformations. An analytic expression for the hopping term is then used to determine the parameters of the system, allowing for the arbitrary single-qubit transformation of an electron as a result of its transition around the ring. The paper is organized as follows: the model describing the electron on the ring is introduced in Section 2 and the Bloch states on the ring are derived by analytical solving the Schr\"{o}dinger equation in Section 3. In Section 4 the Wannier states on the ring are introduced  and in Section 5 transformed into spin Wannier basis. These finally enables the analysis of qubit transformations, which is done in Section 6, and Section 7 is devoted to conclusions.

\section{Model}
The Hamiltonian, governing the electron on the ring in presence of Rashba coupling and external potential, is given by\cite{Meijer2002}
\begin{equation}
\label{eq:H0}
H =\epsilon \left( i \partial_\varphi + \phi_m \right)^2  -\alpha \epsilon \left[ \sigma_\rho(\varphi) \left(i \partial_\varphi +\phi_m\right) + \frac{i}{2} \sigma_\varphi(\varphi) \right] + V(\varphi),
\end{equation}
with parameters
\begin{equation}
\label{eq:1_1_23a}
\frac{\hbar^2}{2 m R^2} \equiv \epsilon, \quad \frac{2 m R \alpha_R}{\hbar} \equiv \alpha, \quad  \frac{\phi}{\phi_0} \equiv \phi_m,
\end{equation}
where periodic angular coordinate $\varphi \in \left[ 0, 2 \pi \right]$ describes the position of the electron. $R$ denotes the ring radius, $m$ the electron effective mass in a semiconductor, $\alpha_R$ the Rashba coupling, $\phi$ magnetic flux through the ring and $\phi_0$ magnetic flux quantum. Pauli operators in rotated spin frame are defined as 
\begin{equation}
\label{eq:1_1_3}
\eqalign{
\sigma_\rho\left( \varphi \right) &=\phantom{-} \sigma_x \cos \varphi + \sigma_y \sin \varphi,  \nonumber \\
\sigma_\varphi\left( \varphi \right) &=  -\sigma_x \sin \varphi + \sigma_y \cos \varphi,
}
\end{equation}
where $\sigma_{x,y}$ are ordinary Pauli matrices.
In our model, $V(\varphi)$ is a periodic potential with the period $ \varphi_a = 2 \pi/N$, described as a sum of $N$ potential wells $W(\varphi)$, shifted to have minima at $\varphi = n \varphi_a$,
\begin{equation}
\label{eq:pot_W}
V(\varphi) = \sum_{n=1}^N a_n W(\varphi - n \varphi_a).
\end{equation}
Coefficients $a_n$ describe the depth of the potential at each site and can be varied externally by the voltage applied to each gate. These allow the transfer of the electron around the ring. To keep the electron located at site $n$, the depth of the potential well on this site, $a_n$, should be set to sufficiently large value while all other coefficients should be set to $0$. To transfer the electron to the neighboring site, $n\pm1$, coefficients $a_{n \pm 1}$ should be increased, respectively, while $a_n$ is simultaneously set to $0$.

\section{Schr\"odinger equation}
\label{sec:Sec2}
The main goal of this paper is to calculate analytically how the spin orientation of the electron changes during this process. As we show later, this information is encoded in the hopping terms for an electron between gate positions, which can be extracted from Bloch states $\psi_{js}(\varphi)$ with their energies $E_{js}$, obtained for the case of equal binding potentials on all gate sites on the ring, $a_n = 1$. The Schr\"odinger equation for Bloch states is
\begin{equation}
\label{eq:SchEqu}
H \psi_{js}(\varphi) = E_{js} \psi_{js}(\varphi),
\end{equation}
where half-integer index $j$ is used to denote the rotation symmetry of the wavefunction and $s = \pm \frac{1}{2}$ is a pseudo-spin index. The symmetry properties of ring Hamiltonian equation \eqref{eq:H0} lead to an ansatz for Bloch function, derived in \ref{app:BlochStates},
\begin{equation}
\label{eq:psijs_ans}
\psi_{js}(\varphi) = e^{i \left( j -\frac{1}{2} \sigma_z\right)} u_j(\varphi) \chi^*_s,
\end{equation}
with $u_{j}(\varphi)$ being periodic function of $\varphi$, $u_{j}(\varphi+ \varphi_a) = u_{j}(\varphi)$. To find an exact form of periodic function $u_{j}(\varphi)$ and spinor $\chi_s^*$ for the case of Rashba Hamiltonian equation \eqref{eq:H0}, we transform it with a set of unitary transformations, given in Ref.~\cite{Kregar2016b}
\begin{equation}
\label{eq:Utrans}
U = U_\alpha U_z U_\phi = e^{-i \frac{\varphi}{2} \vec{\alpha} \cdot \vec{\sigma}} e^{i \frac{\varphi}{2} \sigma_z} e^{-i \phi_m \varphi},
\end{equation}
where $\vec{\alpha} = \left( -\alpha, 0, 1 \right)$ is an effective Rashba field and $\vec{\sigma} = \left( \sigma_x, \sigma_y, \sigma_z \right)$ is the vector of standard Pauli operators. The transformation does not affect the periodic potential $V(\varphi)$ and the resulting Hamiltonian is independent of spin
\begin{equation}
\label{eq:Hcrta}
H' = U H U^\dagger = - \epsilon \partial_\varphi^2 + V(\varphi) + E_{SO},
\end{equation}
with spin-orbit energy $E_{SO} = - \frac{1}{4} \epsilon \alpha^2$. As explained in \ref{app:BlochStates}, this spin-independent form allows one to seek the Bloch states in a manner very similar to the case of electron on a one-dimensional straight wire with periodic potential, i.e. using the ansatz 
\begin{equation}
\psi'_{ks} (\varphi) = e^{i k \varphi} u'_k(\varphi) \chi^*_s.
\end{equation}
The Bloch states of original Hamiltonian are then obtained by inverse transformation
\begin{equation}
\label{eq:Udag_psi}
 \psi_{ks} (\varphi) = U^\dagger \psi'_{ks}(\varphi).
\end{equation}
The values of $k$ and eigenspinors $\chi^*_s$ are determined by applying non-trivial periodic boundary conditions, $\psi_{ks}(\varphi) = \psi_{ks}(\varphi + 2\pi)$, resulting in \cite{Kregar2016b}
\begin{equation}
\label{eq:BC}
\chi_s^* = - e^{2 \pi i \left( \phi_m + k \right) } e^{i \frac{\varphi}{2} \vec{\alpha}\cdot \vec{\sigma}} \chi_s^*.
\end{equation}
The eigenproblem has two solutions, one for each pseudo-spin state $s$. Both can be compactly written as a spin transformation of standard basis spinors, quantized along the $z$-axis, denoted $\chi_s$, using an operator of spin rotation around the $y$-axis, $U_y(\vartheta_\alpha) = \exp \left( - i \frac{\vartheta_\alpha}{2} \sigma_y \right)$,
\begin{equation}
\label{eq:chi_s_zv}
\chi_s^* = U_y(\vartheta_\alpha) \chi_s, \quad \tan \vartheta_\alpha = -\alpha.
\end{equation}
Applied to boundary conditions equation \eqref{eq:BC}, the spinors equation \eqref{eq:chi_s_zv} determine the allowed values of $k$, which also depend on pseudo-spin $s$.
\begin{equation}
k_s = j -\phi_m -s \phi_\alpha, \quad j+\frac{1}{2} \in \mathbb{Z}, \quad \phi_\alpha = \sqrt{1+\alpha^2}.
\end{equation}
When applied in ansatz equation \eqref{eq:Udag_psi}, these results lead to the Bloch functions of the Rashba ring Hamiltonian equation \eqref{eq:H0} being expressed analytically as
\begin{equation}
\label{eq:BlochStatePsijs}
\psi_{js} (\varphi) = e^{i j \varphi} u_{j s}(\varphi) U_z^\dagger(\varphi) U_y^\dagger(\vartheta_\alpha) \chi_s.
\end{equation}
What is important is that the periodic part of the Bloch function $u_{j s}(\varphi)$ can be directly related to the function $u'_k(\varphi)$ for the case of one-dimensional system,
\begin{equation}
u_{j s}(\varphi) = u'_{k}(\varphi),
\end{equation}
by substituting $k\rightarrow j - \phi_m - s\phi_\alpha$, given that the periodic part of Hamiltonian $V(\varphi)$ is the same in both cases. Note that when exponent $e^{i j \varphi}$ is combined with spin rotation $U_z^\dagger(\varphi)$, the result is indeed compatible ansatz equation \eqref{eq:psijs_ans}, derived in \ref{app:BlochStates}.

The energy of one-dimensional Bloch state in the limit of strong periodic potential (tight-binding limit) is parametrised as $E_{k} = E_0 -2 t_0 \cos \left( k \varphi_a  \right)$, with mean band energy $E_0$ and bandwidth $4 t_0$ determined by detailed shape of the potential \cite{Kittel2004}. The transformation between the one-dimensional and the ring Hamiltonian allows the energy of the electron on the ring to be obtained by a simple substitution introduced above, $k \rightarrow j - \phi_m - s\phi_\alpha$, into the expression for $E_{k}$, resulting in energy depending on both angular momentum $j$ and pseudo-spin $s$,
\begin{equation}
\label{eq:BlochEnjs}
E_{js} = E_0  + E_{SO}- 2 t_0 \cos \left( \varphi_a  \left[ j - \phi_m - s\phi_\alpha \right]  \right).
\end{equation}

Since both Bloch states of equation \eqref{eq:BlochStatePsijs} and energies equation \eqref{eq:BlochEnjs} on the ring closely resemble their one-dimensional counterparts, their transformation to Wannier states and their corresponding Hamiltonian is obtained by a simple transformation, presented in the next section.

\section{Wannier states}
As explained in the Introduction, the spin transformations, accompanying the electron's transition between sites on a ring, will be expressed in terms of nearest-neighbor hopping terms. These are obtained by the Fourier transformation of Bloch states into the basis of localized Wannier functions \cite{Kittel2004},
\begin{equation}
\label{eq:wann_def}
\eqalign{
\phi_{ns}(\varphi) = & \frac{1}{\sqrt{N}} \sum_{j=\frac{1}{2}}^{N-\frac{1}{2}} e^{- i n \left(j - \frac{1}{2} \right)\varphi_a} \psi_{js} (\varphi) = \nonumber \\
&= e^{i \frac{\varphi}{2}} w_{ns}(\varphi) U_z^\dagger(\varphi) U_y^\dagger(\vartheta_\alpha) \chi_s.}
\end{equation}
Note that since summation is taken over half-integer $j$ values, the phase coefficients $e^{- i n \left(j - \frac{1}{2} \right)\varphi_a}$ are such that $j-\frac{1}{2}$ is an integer, as is usual for the Fourier transformation. We used the fact that transformations $U_z^\dagger$ and $U_y^\dagger$ do not depend on $s$, so the envelope function $w_{n s}(\varphi)$, describing the charge density of the wavefunction, is a Fourier transformation of $u_{js}(\varphi)$,
\begin{equation}
\label{eq:4_wndef}
w_{n s}(\varphi)= \frac{1}{\sqrt{N}} \sum_{j} e^{i  \left(j - \frac{1}{2} \right)\left( \varphi - n \varphi_a \right)} u_{j s}(\varphi).
\end{equation}

\begin{figure}[ht]
\centerline{\includegraphics[scale=1]{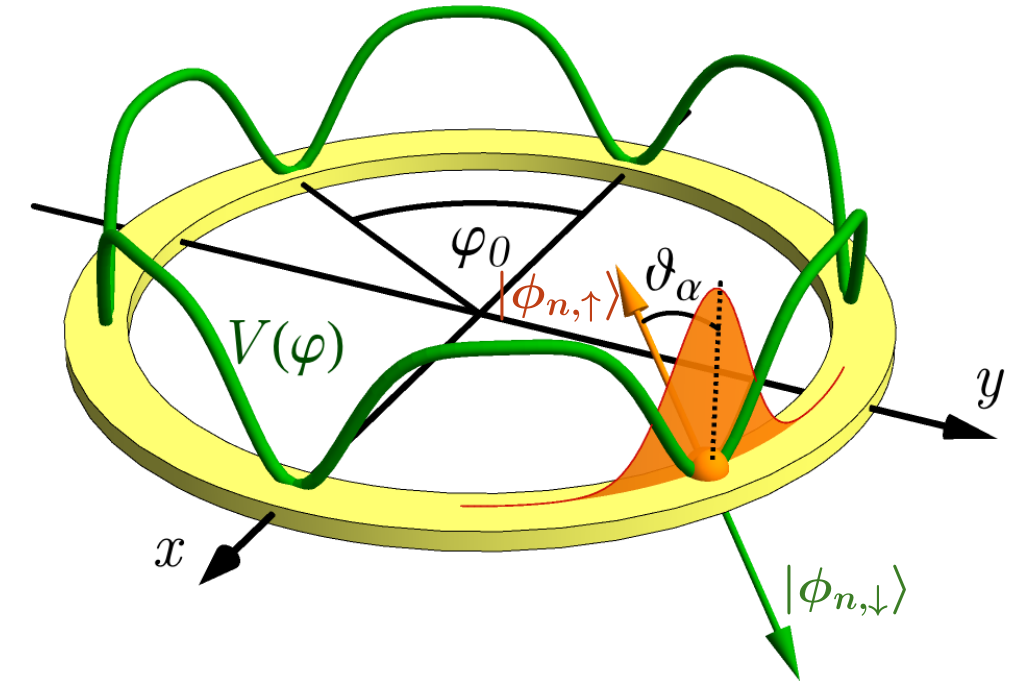}}
\caption[Ring-schematic]{Schematic representation of the Wannier state $\left| \phi_{n s} \right\rangle$ as an electron, localized at a minimum of periodic potential, with tilted spin.}
\label{fig:wann_spin}
\end{figure}

The expectation value of spin of the Wannier function
\begin{equation}
\eqalign{
\left\langle \vec{s} \right\rangle_{ns} &=  \frac{\hbar}{2} \int_{-\pi}^\pi \left| w_{ns}(\varphi)\right|^2 \chi_s^\dagger U_y(\vartheta_\alpha)  U_z(\varphi) \vec{\sigma} U_z^\dagger(\varphi) U_y^\dagger(\vartheta_\alpha) \chi_s {\rm d}\varphi = \nonumber \\
&= \hbar s \int_{-\pi}^\pi \left| w_{ns}(\varphi)\right|^2  \left( \sin \vartheta_\alpha \cos \varphi ,\sin \vartheta_\alpha \sin \varphi , \cos \vartheta_\alpha \right) {\rm d}\varphi.
}
\end{equation}
is mostly determined by spin rotations $U_z(\varphi)$ and $U_y(\vartheta_\alpha)$. If the periodic potential is strong, functions $w_{ns}(\varphi)$ are strongly localized around positions $\varphi = n \varphi_a$ and expectation values of spin can reliably be approximated by 
\begin{equation}
\label{eq:spinVal}
\left\langle \vec{s} \right\rangle_{ns} = \hbar s \left( \sin \vartheta_\alpha \cos \left( n \varphi_a \right) ,\sin \vartheta_\alpha \sin \left( n \varphi_a \right) , \cos \vartheta_\alpha \right).
\end{equation}
This leads to very intuitive interpretation of the Wannier states and their spin properties. The electron in the Wannier state $\left| \phi_{ns} \right\rangle$ is localized around the position $n\varphi_a$ with spin tilted from $z$ direction towards the centre of the ring for $s=1/2$ and from $-z$ direction away from the centre for $s=-1/2$, as shown in figure \ref{fig:wann_spin}.

The matrix elements of Hamiltonian $H$ in the Wannier basis $H_{m n s s'} = \left\langle \phi_{m s} \right| H \left| \phi_{n s'} \right\rangle$ are obtained as the Fourier transformation of energy $E_{js}$,
\begin{equation}
H_{m n s s'} = \frac{1}{N} \delta_{s s'} \sum_{j=\frac{1}{2}}^{N-\frac{1}{2}}  e^{i \left(j -\frac{1}{2} \right) \left( m - n\right) \varphi_a} E_{ j s}.
\end{equation}
Since $j$ only appears in cosine terms in $E_{js}$, the transformed Hamiltonian can be exactly evaluated,
\begin{equation}
\label{eq:Hmns}
H_{m n s}  = E_0 + E_{SO} -t_s \delta_{m, n+1} -t_s^* \delta_{m, n-1},
\end{equation}
with pseudo-spin dependent hopping term
\begin{equation}
\label{eq:ts}
t_s= t_0 e^{i \varphi_a \left( \frac{1}{2} - \phi_m - s\phi_\alpha  \right)}.
\end{equation}
The Hamiltonian $H$ in the basis of the Wannier states therefore correspond to a tight-binding model with spin dependent hopping term $t_s$,
\begin{equation}
\label{eq:Hhopping}
H = E_0 + E_{SO} - \sum_{n s} \left( t_s  \left| \phi_{n+1,s} \right\rangle \left\langle \phi_{n s} \right| + t_s^* \left| \phi_{n-1,s} \right\rangle \left\langle \phi_{n s} \right| \right).
\end{equation}

\section{Spin Wannier basis}
Application of hopping terms $t_s$ in equation~\eqref{eq:Hhopping}, although simple, is not the best way to study spin transformations. Since $t_s$ couples states $\left| \phi_{n s} \right\rangle$ with a non-trivial spin properties equation \eqref{eq:spinVal}, the interpretation of the effect of hopping on electron's spin orientation is more complicated. This issue is tackled here by introducing a  basis of localized states with uniform spin orientation, as follows.

Since the spin properties of Wannier functions depend on the strength of the Rashba coupling $\left| \phi_{n s} \right\rangle$, these states are not the best choice for the analysis of spin transformations of the electron. It is more convenient to construct a new basis states as a local superposition of Wannier states at the same site $n$, so-called spin Wannier basis,
\begin{equation}
\tilde{\phi}_{ns}(\varphi) = \sum_{s'} c_{n s' s} \phi_{ns'}(\varphi).
\end{equation} 
with spin properties {\it independent} of spin-orbit coupling, resembling pure spin states. We construct these states in a way that their expectation values of spin are as close as possible to the values for pure spin states,
\begin{equation}
\label{eq:sns_sWann}
\left\langle \vec{s} \right\rangle_{ns} = \left\langle \tilde{\phi}_{n s} \right| \vec{s} \left| \tilde{\phi}_{n s} \right\rangle \approx \hbar s \vec{e}_z,
\end{equation}
as explained in \ref{app:WannSpinBase}. To emphasize that this basis resembles pure spin states, we sometimes use arrows $\uparrow$ and $
\downarrow$ as the pseudo-spin index $s$ instead of $\pm \frac{1}{2}$, respectively. The coefficients of linear superposition of such states
\begin{equation}
\left| \psi \right\rangle = \cos \left( \frac{\theta}{2} \right) \left| \tilde{\phi}_{n \uparrow} \right\rangle + e^{i \chi} \sin \left( \frac{\theta}{2} \right)\left| \tilde{\phi}_{n \downarrow} \right\rangle
\end{equation}
can then be directly related to the direction the vector of spin expectation values on the Bloch sphere, $\theta$ and $\chi$
\begin{equation}
\left \langle \psi \right| \vec{s} \left| \psi \right\rangle \approx \frac{\hbar}{2} \left( \sin \theta \cos \chi, \sin \theta \sin \chi, \cos \theta \right),
\end{equation}
which significantly simplifies the analysis of spin transformations and makes the states $\left| \tilde{\phi}_{n s} \right\rangle$ a suitable qubit basis.

The coefficients $c_{n s s'}$ are determined by projecting the original Wannier states to the basis of pure spin states, as show in \ref{app:WannSpinBase}. In the limit of strongly localized states $\left| \phi_{n s} \right\rangle$, the coefficients simplify to
\begin{equation}
c_{n s s'} = e^{-i n \frac{\varphi_a}{2}} \mathcal{U}_{n s s'},
\end{equation}
where the matrix $\mathcal{U}$ can be expressed with spin rotations $U_z(\varphi)$ and $U_y(\vartheta_\alpha)$, introduced in the Hamiltonian transformation equation \eqref{eq:Utrans},
\begin{equation}
\label{eq:Unss}
\mathcal{U}_{n s s'} = \chi^\dagger_s U_y(\vartheta_\alpha) U_z(n\varphi_a) \chi_{s'}.
\end{equation} 
Even though this result is not exact, these coefficients  represent a good approximation of pure spin states even for the case of shallow potential wells, as is demonstrated numerically in figure \ref{fig:spin_mod_wann} in \ref{app:WannSpinBase}. 

Since the spin Wannier state $\left| \tilde{\phi}_{n s} \right\rangle$ is a local superposition of original Wannier states $\left| \phi_{n s} \right\rangle$, with the same $n$, the Hamiltonian in this basis will still have a form of nearest neighbor hopping, but with coupling terms $\tilde{t}_{n s s'}$ being position-dependent and also mixing the pseudo-spin states,
\begin{equation}
\label{eq:HamSWann}
H = E_0 + E_{SO} - \sum_{n s s'} \left( \tilde{t}_{n s s'}^+   \left| \tilde{\phi}_{n+1,s} \right\rangle \left\langle \tilde{\phi}_{n s'} \right| + \tilde{t}_{n s s'}^- \left| \tilde{\phi}_{n-1,s} \right\rangle \left\langle \tilde{\phi}_{n s'} \right| \right).
\end{equation}
Hopping terms $\tilde{t}_{n s s'}^\pm$ are calculated by transforming $t_s$ equation \eqref{eq:ts} with the matrix $\mathcal{U}_n$ equation \eqref{eq:Unss}
\begin{equation}
\label{eq:hoppingt}
\eqalign{
\tilde{t}_{n s s'}^\pm&= t_0 e^{\pm i \varphi_a \phi_m} \chi_s^\dagger U_z^\dagger(n \varphi_a \pm \varphi_a) U_\alpha^\dagger(\pm \varphi_a) U_z(n \varphi_a) \chi_{s'} = \\
 &=  t_0 e^{\pm i \varphi_a \phi_m} \left(
\begin{array}{cc}
 e^{\mp \frac{1}{2} i  \varphi_a} (c_\phi+i c_\alpha s_\phi) & i
   s_\alpha s_\phi e^{- i \varphi_a \left( n \pm \frac{1}{2} \right)} \\
 i s_\alpha s_\phi e^{i \varphi_a \left(n \pm \frac{1}{2} \right)} & e^{ \pm \frac{1}{2}\varphi_a} (c_\phi-i c_\alpha s_\phi)
\end{array}
\right)_{ss'},}
\end{equation}
where
\begin{equation}
\eqalign{
\label{eq:sc_def}
s_\alpha &= \sin \vartheta_\alpha \quad s_\phi = \sin\left(\frac{\varphi_a}{2} \phi_\alpha \right), \nonumber \\
c_\alpha &= \cos \vartheta_\alpha \quad c_\phi = \cos\left(\frac{\varphi_a}{2} \phi_\alpha \right).
}
\end{equation}
Although not obvious at first glance, the Hamiltonian equation \eqref{eq:HamSWann} is Hermitian when applied to the basis of states $\left| \tilde{\phi}_{n s} \right\rangle$ with appropriate periodic boundary conditions on a ring.
 
The hopping terms $\tilde{t}_{n s s'}^\pm$ are quite complex, but still expressed in analytical form, comprising three spin-rotation matrices. In contrast to $t_s$, describing the transformation of pseudo-spin states $\left| \phi_{n s} \right\rangle$ with relatively complex spin properties (see Fig. \ref{fig:wann_spin}), the interpretation of terms $\tilde{t}_{n s s'}^\pm$ is much more direct, describing real spin rotations, expressed in spin Wannier basis $\left| \tilde{\phi}_{n s} \right\rangle$. Consequently, this allows a much simpler analysis of spin rotations, accompanying electrons movement between voltage gate sites, and also a construction of general single-qubit transformations. This will be further explored in the next section by the introduction of suitable qubit basis and demonstration of system capabilities in performing controlled qubit transformations.

\section{Qubit transformations}
We define qubit basis as Wannier pseudo-spin pair on the site $n=0$,
\begin{equation}
\left| 0 \right\rangle = \left| \tilde{\phi}_{0 \uparrow} \right\rangle, \quad \left| 1 \right\rangle = \left| \tilde{\phi}_{0 \downarrow} \right\rangle.
\end{equation}
We also define the Bloch sphere, corresponding to this basis, defined by polar and azimuthal angles $\Theta$ and $\Phi$, which correspond to the qubit state
\begin{equation}
\label{eq:Qubit}
\left| \psi_{Q} \right\rangle = \cos \left( \frac{\Theta}{2}\right) \left| 0 \right\rangle +  e^{ i \Phi} \sin \left(\frac{\Theta}{2}\right)\left| 1 \right\rangle.
\end{equation}

Single qubit transformation is achieved by transferring the electron around the ring by controlled changes of gate potentials at different sites. To transfer the electron from one site to its neighboring site, we slowly decrease the depth of potential well on the first site and increase the depth of the potential on the site onto which we want to transfer the electron. Such charge transfer has already been demonstrated experimentally for $N=4$ sites \cite{Thalineau2012}. From mathematical perspective, this results in a Landau-Zenner-like transition of the electron from the superposition of spin Wannier states $\left| \tilde{\phi}_{n s} \right\rangle$ on the initial site to the superposition of spin Wannier states $\left| \tilde{\phi}_{n+1, s} \right\rangle$ on the final site, as analysed in \ref{app:Zenner}. 

As in the case of the Landau-Zenner transition, the probability of finding the electron on the initial site will drop to zero only in a case of slow change of the local potential. Even in this limit, however, the resulting transition is not trivial, since the coefficients of the spin superposition change during the transition. The change is described by the hopping term for spin Wannier basis equation \eqref{eq:hoppingt}. If the electron is initially in a state on-site $n$
\begin{equation}
\label{eq:psiInit}
\left| \psi_{init} \right\rangle = \sum_s c_s \left| \tilde{\phi}_{n s} \right \rangle,
\end{equation}
the Landau-Zenner transition from site $n$ to $n+1$, denoted $T_{n \rightarrow n+1 }$, will result in the final state (see \ref{app:Zenner})
\begin{equation}
\label{eq:ZennTrans}
\left| \psi_{fin} \right\rangle = T_{n \rightarrow n+1} \left| \psi_{init} \right\rangle = \sum_s d_s \left| \tilde{\phi}_{n+1,s} \right \rangle
\end{equation}
with new coefficients $d_s$ calculated from the hopping term $\tilde{t}_{n s s'}^+$:
\begin{equation}
\label{eq:coefds}
d_s = \sum_{s'} \frac{1}{t_0} \tilde{t}_{n s s'}^+ c_{s'}.
\end{equation}
Note that the $2\times 2$ matrix 
\begin{equation}
\tilde{\mathcal{U}}_{n s s'}^+(\alpha) = \tilde{t}_{n s s'}^+/t_0
\end{equation}
is unitary, as is seen from equation \eqref{eq:hoppingt}, which means that each transition can be seen as a rotation on the Bloch sphere. Note that the transformation of coefficients depends on the strength of the Rashba coupling $\alpha$, determining the axis of spin rotation $U_\alpha^\dagger$ in hopping term equation \eqref{eq:hoppingt}.

The sequence of Landau-Zenner transitions equation \eqref{eq:ZennTrans} between neighboring sites can bring the electron around the entire ring, resulting in the final state being a superposition of the same spin Wannier states $\left| \tilde{\phi}_{N+n, s} \right \rangle = \left| \tilde{\phi}_{n s} \right \rangle$ as the initial state
\begin{equation}
\label{eq:psiFin}
\left| \psi_{fin} \right\rangle = \sum_s d_s \left| \tilde{\phi}_{n s} \right \rangle.
\end{equation}
The coefficients describing the final state are calculated as
\begin{equation}
d_s = \sum_{s'} \mathcal{U}_{full,ss'} c_{s'},
\end{equation}
with transformation $\mathcal{U}_{full}$ being a product of spin transformation for each transition between neighboring sites.
\begin{equation}
\label{eq:Ufull}
\mathcal{U}_{full} = \tilde{\mathcal{U}}_{N-1}^+(\alpha_{N-1}) \cdot\tilde{\mathcal{U}}_{N-2}^+(\alpha_{N-2})  \cdots\tilde{\mathcal{U}}_{1}^+(\alpha_{1})  \tilde{\mathcal{U}}_{0}^+(\alpha_{0}).
\end{equation}
Since the Rashba coupling $\alpha_i$ can be adjusted between two consequential Landau-Zenner transitions, this gives a wide range of parameters that can be tuned to achieve desired qubit transformation.

Using the definition of $\tilde{t}_{n s s'}^+$ equation \eqref{eq:hoppingt} and allowing $m$ revolutions of the electron around the ring with $N$ sites, the qubit transformation $\tilde{\mathcal{U}}_{full}$ can be written in a simplified manner (the rotations $U_z^\dagger$ cancel out) using only spin transformations $U_\alpha^\dagger$,
\begin{equation}
\label{eq:Ufullss}
\mathcal{U}_{full,s s'} = (-1)^m \chi_s^\dagger U_{\alpha_{m \times N-1}}^\dagger(\varphi_a) \cdots U_{\alpha_{0}}^\dagger(\varphi_a) \chi_{s'},
\end{equation}
where each factor corresponds to an electron's transition between sites at the Rashba coupling strength $\alpha_i$, with $i=0, 1,  ... , m\times N - 1$. Note that the phase factor $(-1)^m$ (arising from $U_z^\dagger(2\pi) = -1$) depends on the number of electron's revolutions around the ring, but does not physically affect the spin transformation. 

By using the qubit states $\left| \psi_Q \right\rangle$ equation \eqref{eq:Qubit} as the initial state $\left| \psi_{init} \right\rangle$ equation \eqref{eq:psiInit}, the final state $\left| \psi_{fin} \right\rangle$ equation \eqref{eq:psiFin} is also a qubit and the transformation equation \eqref{eq:Ufullss} therefore represents a controlled qubit transformations. It is instructive to see it as a combination of rotations on the Bloch sphere, spanned by qubit basis, where each transition of the electron, described by transformation $U_{\alpha_n}^\dagger(\varphi_a) = \exp \left( \frac{1}{2} i \varphi_a \vec{\alpha}_n\cdot \vec{\sigma} \right)$, causes a rotation around axis $\vec{\alpha}_n = \left( -\alpha_n, 0, 1\right)$ by the angle $\chi_n = \varphi_a \sqrt{1 + \alpha_n^2}$. The result is very similar to the one found in Ref.~\cite{Kregar2016}, but in the present case, the shifts in electrons position $\varphi_a$ are fixed and the strength of Rashba coupling during each transition can be tuned. Note that the present case is much closer to the model of a possible realistic device, where the electron would be transferred between the potential minima, defined by potential gates at fixed positions.

To verify that the described procedure can really be used to realize a qubit gate, we performed comprehensive numerical calculations, similar as in Ref.~\cite{Kregar2016}. We describe the total qubit transformation with angles on the Bloch sphere $\Theta$ and $\Phi$, corresponding to the final qubit state, obtained from initial state $\left| 0 \right\rangle$ by applying transformation $\mathcal{U}_{full}$. The transformation is determined by a set of Rashba parameter values $\alpha_i$, which can take values between intrinsic, non-amplified value $\alpha_{in}$ and amplified value $\alpha_{max} = K_{\alpha} \alpha_{in}$ with $K_{\alpha}$ depending on the material used. As in Ref.~\cite{Kregar2016} we choose the ring size $R$ in such a way that $\alpha_{in} = 1/\sqrt{K_{\alpha}}$ and $\alpha_{max} = \sqrt{K_{\alpha}}$ (see equation \eqref{eq:H0}), providing the maximal angle between rotation axis corresponding to these two values of $\alpha$. For each number of sites on a ring $N$, the number of revolutions $m$ and maximal amplification factor $K_{\alpha}$, parameters that are determined by device architecture and material, a set of numbers $\left[ \alpha_0, ..., \alpha_{N\times m-1} \right]$ determines the qubit transformation, parametrized by $\Theta$ and $\Phi$. If we can for each pair of $\Theta$ and $\Phi$ find a set $\left[ \alpha_0, ..., \alpha_{N\times m-1} \right]$, this means that any qubit transformation can be achieved.

\begin{figure}[ht]
\centerline{\includegraphics[scale=1]{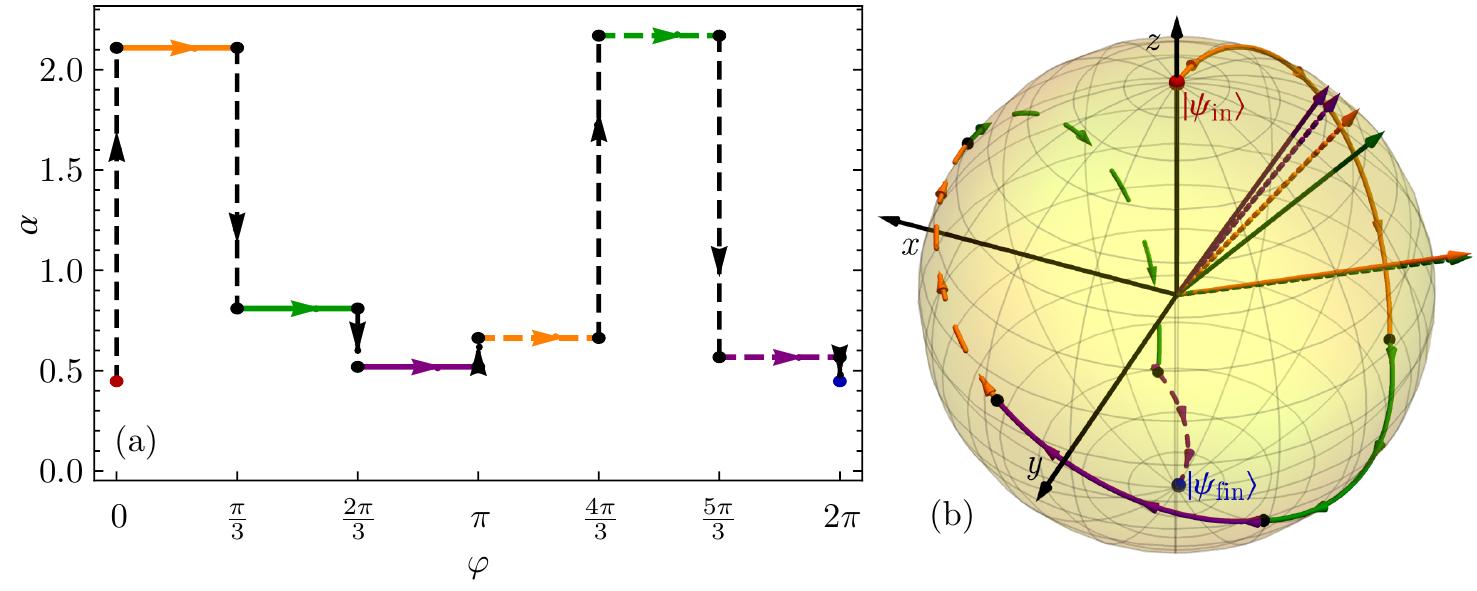}}
\caption[Z-transformation]{An example of qubit Z-gate transformation. (a) Movement of the system in parametric space with coordinates being electron's position $\varphi$ and Rashba coupling $\alpha$, transforming the initial state $\left| \psi_{in} \right\rangle = \left| 0 \right\rangle$ (red dot) to the final state $\left| \psi_{fin} \right\rangle = \left| 1 \right\rangle$ (blue dot). Before each shift of electron using Landau-Zenner transition, the value of Rashba coupling is adjusted to the appropriate value, calculated using Monte-Carlo simulation. (b) Resulting spin transformations are represented as a rotations around axes, determined by the Rashba coupling. Orange, green and purple solid and dashed lines on (a) correspond to the rotational axes and spin rotation paths on (b).}
\label{fig:Ztrans}
\end{figure}

As an example of spin rotation, we performed the Z-gate qubit transformation, corresponding to $\Theta = \pi$ and arbitrary $\Phi$. This transformation can be realized on a ring with $N=6$ sites with $m=1$ revolution of the electron around the ring and Rashba amplification factor $K_{\alpha}=5$. The transformation is schematically presented in figure \ref{fig:Ztrans}. figure \ref{fig:Ztrans}(a) shows how values of the Rashba coupling need to be changed between the shifts of electron position. On figure \ref{fig:Ztrans}(b) the rotations of electron spin is schematically presented on the Bloch sphere with arrows representing the rotation axis of each spin rotation, with colours and dashing corresponding to the ones in figure \ref{fig:Ztrans}(a). Although this representation is very instructive, note that only the initial (red dot) and final (blue dot) state on the Bloch sphere correspond to qubit states, defined as being located at site $n=0$. The intermediate points on Bloch sphere are defined in a space, corresponding to the rotation $\mathcal{U}_{full}$ and can be related to actual physical states only if the full rotation $\mathcal{U}_{full}$ is decomposed back into single-transition rotations $\tilde{\mathcal{U}}_{n s s'}^+(\alpha_n)$ equation \eqref{eq:Ufull} and the intermediate results are expressed in spin Wannier basis $\left| \tilde{\phi}_{ns}\right\rangle$.

\begin{figure}[ht]
\centerline{\includegraphics[scale=1]{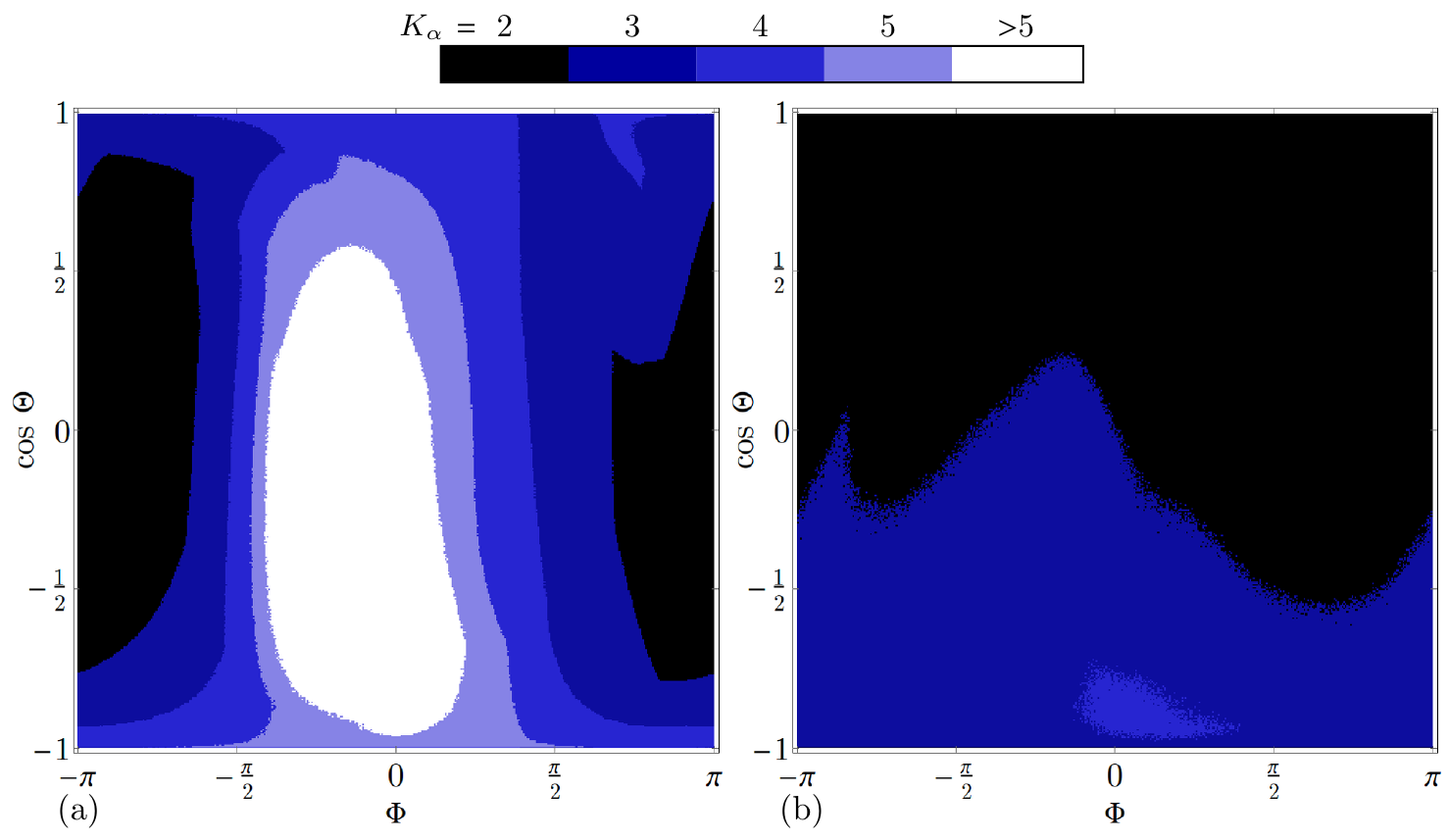}}
\caption[Bloch sphere coverage]{Areas of Bloch sphere covered at different values of Rashba amplification factor $K_{\alpha}$ for a ring with $N=6$ sites. Panel (a) shows results for $m=1$ revolution of electron around the ring and (b) for $m=2$ revolutions. For $m=1$ a part of the sphere remains uncovered at $K_{\alpha}=5$ while for $m=2$, all the sphere is covered even at $K_{\alpha}=4$.}
\label{fig:cont}
\end{figure}

To determine which parts of the Bloch sphere can be covered at specific choice of $N$, $m$ and $K_{\alpha}$, the Monte-Carlo simulation is used. $N_{MC} = 3 \times 10^{11}$ sets $\left[ \alpha_0, ..., \alpha_{N\times m-1} \right]$ were randomly generated for each combination of $N$, $m$ and $K_{\alpha}$, each of them resulting in a point ($\Theta$,$\Phi$) on the Bloch sphere. Plotting the points $(\Phi,\cos \Theta)$ in a 2D diagram shows which parts of the Bloch sphere can be covered at chosen values of $N$, $m$ and $K_{\alpha}$. The results of such Monte-Carlo procedure are presented in figure \ref{fig:cont} for $N=6$ sites and various values of $m$ and $K_{\alpha}$. Figure \ref{fig:cont}(a) shows the coverage of the Bloch sphere for $m=1$ electron revolution with black part showing the surface available at the Rashba amplification factor $K_{\alpha}=2$, dark blue at $K_{\alpha}=3$, medium blue at $K_{\alpha}=4$ and light blue at $K_{\alpha}=5$. The qubit transformations, corresponding to white part of Bloch sphere on figure \ref{fig:cont}(a), can only be achieved at amplification factors $K_{\alpha}>5$, which is difficult to obtain in realistic devices. The same diagram for $m=2$ revolutions is presented in figure \ref{fig:cont}(b). We see that in that case, any qubit transformation can be obtained even at lower amplification factor $K_{\alpha}=4$. 

\begin{figure}[ht]
\centerline{\includegraphics[scale=1]{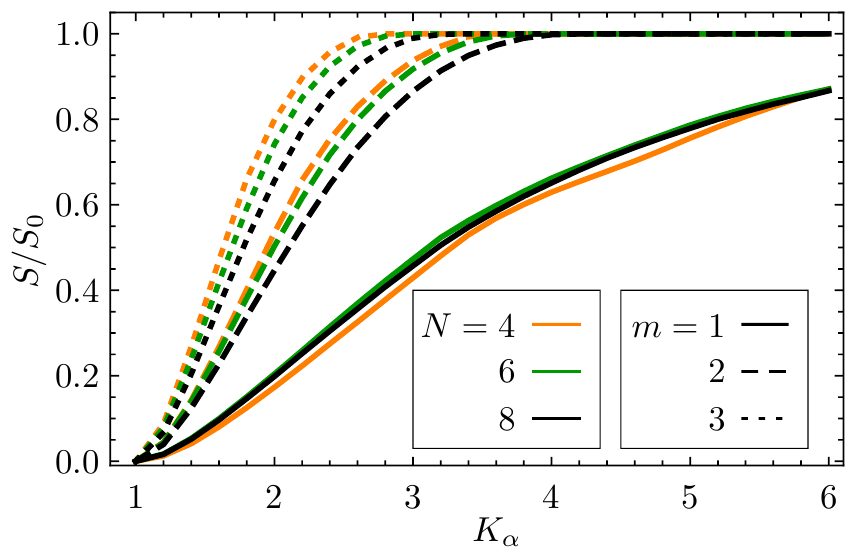}}
\caption[Coverage - k function]{The coverage of Bloch sphere at different numbers of ring sites $N$ and electron's revolutions around the ring $m$ as a function of Rashba amplification factor $K_{\alpha}$. For $m=1$ revolution of electron, the Bloch sphere can not be covered for realistic values of $K_{\alpha}$, while for larger $m$, this can be achieved for $K_{\alpha}$ as low as $4$.}
\label{fig:coverage}
\end{figure}

The dependence of achievable qubit transformations on parameters $N$, $m$ and $K_{\alpha}$ is further explored in figure \ref{fig:coverage}, which shows the percentage of the Bloch sphere that can potentially be covered at specific values of the parameters. We see that the number of revolutions of the electron around the ring is far more important than the number of sites. For $m=2$ revolutions, arbitrary single-qubit rotation can be achieved (fully covered Bloch sphere) with amplification factor $K_{\alpha}\approx4$, while for $N=4$ and $m=3$ the factor $K_{\alpha}$ can be as low as $3$.

\section{Discussion and conclusion}
The results presented here indicate that well-controlled arbitrary transformations of qubits, defined as localized pseudo-spin states of electron on a ring, could be achieved in a quantum ring system where the position of the electron is controlled by a finite number of voltage gates. The efficiency of such an approach, however, depends on several parameters. As discussed in the previous section, the number of shifts of electrons position depends strongly on the maximum amplification factor of the Rashba coupling, achievable in specific material by an external electric field. In simple III-V semiconductor heterostructures, the amplification factors of about $K_{\alpha}=2$ are feasible \cite{Grundler2000,Premasiri2019}, which would lead to a larger number of electron revolutions around the ring. In more exotic systems, for example, InAs nanowires \cite{Liang2012}, a much larger amplification factor of $K_{\alpha}=6$ was measured, however, it is not clear whether such a system is suitable for construction of the quantum ring considered in our study.

The time efficiency of the proposed transformation is to a large extent determined by the size of the ring used. At realistic values of Rashba parameters, the radius of about 100\,nm is required \cite{Kregar2016}, resulting in characteristic energy of electron being about $\epsilon \sim 100\,\mu$V and characteristic time $\tau_0 = \hbar/\epsilon \sim 10^{-11}$\,s. As shown in  \ref{app:Zenner}, the effective Landau-Zenner transitions are achieved at transition times of few tens of characteristic times, which still allows for several thousand electron transitions during spin relaxation time of $100\,\mu$s, typical in semiconductor heterostructures \cite{Bayat2013}.

Note, however, that the Landau-Zenner type transition was chosen in our study due to its simplicity to demonstrate the spin transformations during electrons revolution around the ring. In realistic applications, more efficient and faster ways of electron transport would most likely be applied, which are more demanding for theoretical description but are based on the same phenomena as discussed in this paper. Several other aspects should be taken into account when designing real devices, such as effects of temperature and most importantly the effects of local gate potential, used for the electron transport, on the magnitude of Rashba coupling, which might have an important effect on the spin properties of pseudo-spin states used as qubit basis. Although these effects might change the detailed behaviour of the analyzed system, its ability to performing spin transformations, presented in the paper, would probably not change significantly.


\appendix
\section{Bloch states on Rashba ring}
\label{app:BlochStates}

To find the correct form of Bloch states for an electron on a Rashba ring, described by Hamiltonian equation \eqref{eq:H0}, the symmetry properties of the system are compared to the system of an electron moving in one-dimensional periodic potential, described by the Hamiltonian 
\begin{equation}
H_{1D} = -\frac{\hbar^2}{2 m} \partial_x^2 + V_{1D}(x),
\end{equation}
composed of kinetic energy and periodic potential $V_{1D}(x)$. The Bloch states of such Hamiltonians are typically written as $\psi_{k}(x) = e^{i k x} u_{k}(x)$ with $u_k(x)$ being periodic function \cite{Kittel2004}. The specific form of Bloch functions is a consequence of translation symmetry of periodic potential, which can be described as invariance of the Hamiltonian $H_{1D}$ to the transformation $T(x_0) = \exp\left( -i x_0 p /\hbar \right)$, where $x_0$ is a period of one-dimensional potential $V_{1D}(x)$ and $p = - i\hbar \partial_x$,
\begin{equation}
\label{eq:app_Ham1D}
T(x_0) H_{1D} T^\dagger(x_0)= H_{1D}. 
\end{equation}
Since the Bloch function $\psi_{k}(x)$ should have the same symmetry, the transformation only changes its phase,
\begin{equation}
\label{eq:transTx}
T(x_0)\psi_k(x) = e^{-i k x_0} \psi_k(\varphi).
\end{equation}
The ansatz for Bloch function of an electron in periodic one-dimensional potential is therefore \cite{Kittel2004}
\begin{equation}
\label{eq:app_bloch1D}
\psi_{k} (x) = e^{i k x} u_k(x),
\end{equation}
where $u_k(x)$ is a periodic function of $x$, $u_k(x + x_0) =  u_k(x + x_0)$.

The symmetry of electron states on the Rashba ring, described by Hamiltonian equation \eqref{eq:H0} is a bit more complicated, since it comprises both translation in azimuthal angle by $\varphi_a$ and spin rotation around the $z$-axis by the same angle \cite{Kregar2016b}. The transformation $T_{rot}(\varphi_a)$, corresponding to this symmetry, is generated by the operator
\begin{equation}
J_z = L_z + s_z = \hbar \left( - i \partial_\varphi + \frac{1}{2} \sigma_z \right), \quad T_{rot}(\varphi_a) = e^{-i \frac{\varphi_a J_z}{\hbar}}.
\end{equation}
Similarly to the one-dimensional system, the transformation should only change the phase of the ring Bloch function $\psi_j(\varphi)$,
\begin{equation}
\label{eq:app_Trotpsi}
T_{rot}(\varphi_a) \psi_j(\varphi) = e^{- j \varphi_a} \psi_{j}(\varphi).
\end{equation}
This is indeed true if the ring Bloch function is written as an ansatz, similar to its one-dimensional counterpart equation \eqref{eq:app_bloch1D},
\begin{equation}
\label{eq:app_psijs_ans}
\psi_{js}(\varphi) = e^{i \left( j -\frac{1}{2} \sigma_z\right)} u_j(\varphi) \chi^*_s,
\end{equation}
with function $u_{j}(\varphi)$ being periodic in $\varphi$, $u_{j}(\varphi+ \varphi_a) = u_{j}(\varphi)$. Note that since $T_{rot}$ is a spin operator, the Bloch function is accompanied by some spinor $\chi_s^*$, describing the spin part of the wavefunction, with pseudo-spin index being $s=\pm\frac{1}{2}$. The periodic scalar function $u_j(\varphi)$ depends on half-integer quantum number $j$, which is related to the total angular momentum of the electron.

As shown in Section \ref{sec:Sec2}, the spin-dependent ring Hamiltonian equation \eqref{eq:H0} can be transformed into simplified form using a set of spin transformations $U$ from equation \eqref{eq:Utrans}, $U = U_\alpha U_z U_\phi$.
Since the spin part of the symmetry transformation $T_{rot}$ is already applied to the transformed Hamiltonian $H'$ equation \eqref{eq:Hcrta} in form of a rotation $U_z = \exp \left( i \frac{\varphi}{\hbar} s_z \right)$, $H'$ is invariant under ordinary one-dimensional translation operator, similar to equation \eqref{eq:transTx}, $T(\varphi_a) = \exp\left( -i \varphi_a p_\varphi /\hbar \right)$. This means that $H'$ can for all practical purposes be treated as a Hamiltonian of one-dimensional system $H_{1D}$ equation \eqref{eq:app_Ham1D} and the Bloch states of this transformed Hamiltonian will therefore take a form similar to one-dimensional Bloch state equation \eqref{eq:app_bloch1D}
\begin{equation}
\label{eq:app_psi1}
\psi'_{ks} (\varphi) = e^{i k \varphi} u'_k(\varphi) \chi^*_s,
\end{equation}
but with added spin part $\chi^*_s$. This form differs from equation \eqref{eq:app_psijs_ans} since $k$ in the exponent is a number instead of spin operator. However, once transformed with inverse trasformation $U^\dagger$ equation \eqref{eq:Utrans}, the function takes a form of ansatz equation \eqref{eq:app_psijs_ans} with correct symmetry properties. As for one-dimensional case, the function $u'_k(\varphi)$ is periodic and determined solely by detailed shape of periodic potential $V(\varphi)$ \cite{Kittel2004}, while the spinors $\chi^*_s$ and allowed values of $k$ are determined by the periodic boundary conditions of original Bloch functions, $\psi_{js}(\varphi) = \psi_{js}(\varphi + 2 \pi)$ \cite{Kregar2016b}.

\section{Properties of Wannier spin basis}
\label{app:WannSpinBase}

To calculate the coefficients $c_{n s s'}$, transforming Wannier states $\left| \phi_{n s} \right\rangle$ into spin Wannier basis $\left| \tilde{\phi}_{n s} \right\rangle$, we first construct the basis of pure spin states $\left| \eta_{n s} \right\rangle$, localized at the sites of potential wells,
\begin{equation}
\eta_{n s}(\varphi) = z_n(\varphi) \chi_s
\end{equation}
with orbital part $z_n(\varphi)$ being arbitrary normalized function, strongly localized around coordinate $\varphi = n \varphi_a $, and spin part being pure spinor $\chi_\uparrow$ or $\chi_\downarrow$, quantized along $z$-axis.

We want the spin Wannier basis $\left|\tilde{\phi}_{n s} \right\rangle$ to resemble these states,
\begin{equation}
\label{eq:cns1}
\left| \eta_{n s} \right\rangle \approx \left|\tilde{\phi}_{n s} \right\rangle = \sum_{s'} c_{n s' s} \left|{\phi}_{n s'} \right\rangle,
\end{equation}
so to calculate the coefficients, we simply multiply the equation \eqref{eq:cns1} from the left with Wannier state $\left\langle \phi_{n s''} \right|$,
\begin{equation}
  \left\langle \phi_{n s''} \right. \left| \eta_{n s} \right\rangle \approx c_{n s'' s}.
\end{equation}
When the definition of Wannier states equation \eqref{eq:wann_def} is used in the equation, we get
\begin{equation}
\label{eq:cnss}
c_{n s' s} \approx \int e^{-i \frac{\varphi}{2}} w_{n s'}^*(\varphi) z_n(\varphi) \left[\chi^\dagger_{s'} U_y(\vartheta_\alpha) U_z(\varphi)  \chi_s\right] d\varphi.
\end{equation}
If we assume strong periodic potential, than $w_{n s}(\varphi)$ is narrowly spread around $\varphi = n\varphi_a$. The integration in equation \eqref{eq:cnss} therefore results in elimination of orbital parts of wavefunctions and substitution $\varphi \rightarrow n \varphi_a$ in spin rotations. Also since $w_{n s}(\varphi)$ and $z_{n}(\varphi)$ are generally not orthonormal, the coefficients must be renormalized. This leads to
\begin{equation}
\label{eq:app_coefnss}
c_{n s' s} \equiv  e^{-\frac{n \varphi_a}{2}} \chi^\dagger_{s'} U_y(\vartheta_\alpha) U_z(n \varphi_a)  \chi_s.
\end{equation}
The approximations are rewarded with the fact that the expression is simple and independent of the details of the periodic potential used.

\begin{figure}[ht]
\centerline{\includegraphics[scale=1]{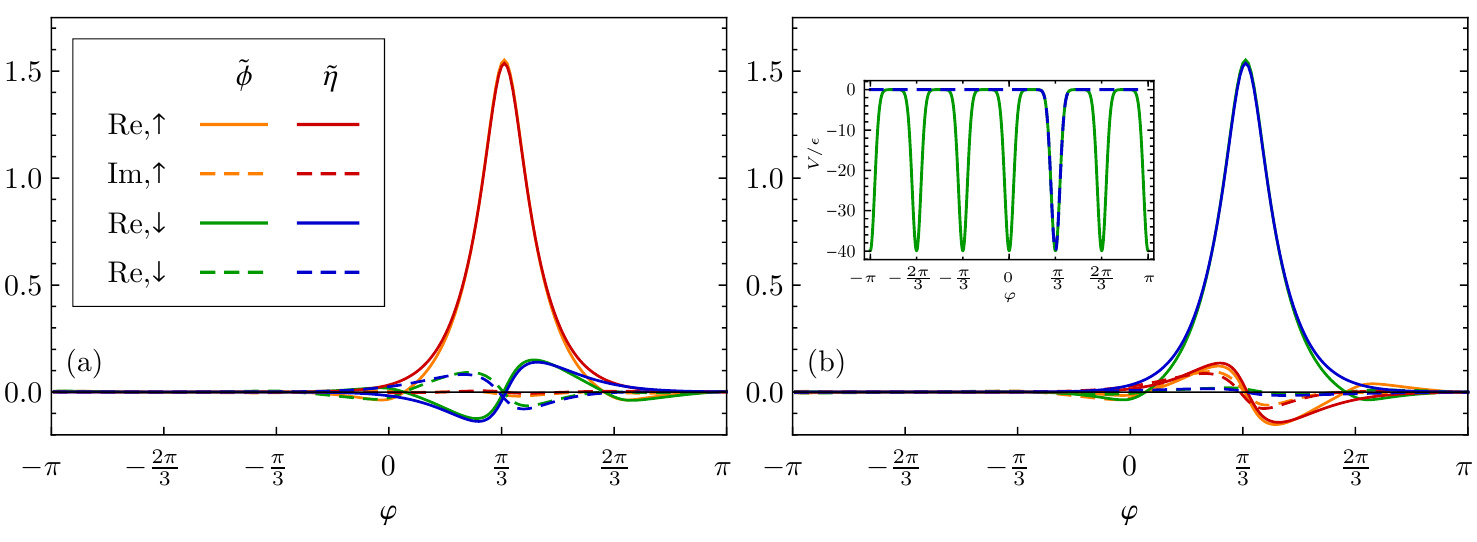}}
\caption[Spin Wannier basis]{Comparison of spin Wannier basis functions $\tilde{\phi}_{ns}(\varphi)$ and bound states $\tilde{\eta}_{ns}(\varphi)$ in Gaussian periodic potential at site $n=1$ with $V_0 =  10 \epsilon$ and $\sigma=0.1$ and Rashba coupling $\alpha = 1.5$, calculated numerically on $N_{num}=240$ sites grid. Panel (a) shows functions $\tilde{\phi}_{n s}(\varphi)$ and   $\tilde{\eta}_{ns}(\varphi)$ for pseudo-spin $s=\frac{1}{2}$ and panel (b) for pseudo-spin $s=-\frac{1}{2}$. Inset figure in (b) shows the periodic potential (green) used to calculate Wannier states $\tilde{\phi}_{n s}$, and single potential well (dashed blue) used to calculate bound states $\tilde{\eta}_{n s}$.}
\label{fig:wann_mod_gauss}
\end{figure}

In order to demonstrate that the coefficients result in a sufficiently good basis functions, we calculate numerically Bloch functions and Wannier functions for the case of periodic potential
\begin{equation}
V(\varphi) = \sum_{n=1}^{N} W(\varphi - n \varphi_a),
\end{equation}
 constructed as a sum of $N=6$ potential wells of Gaussian shape,
\begin{equation}
\label{eq:Wgauss}
W(\varphi) = - \frac{V_0}{\sqrt{2 \pi} \sigma}e^{- \frac{\varphi^2}{2 \sigma^2}}.
\end{equation}
The potential $V(\varphi)$ is characterised by the potential depth $V_0$, corresponding to an integral of the potential over one potential minima, $V_0 = \int_{-\pi}^{\pi} W(\varphi) {\rm d} \varphi$, and its width $\sigma$.

Figure \ref{fig:wann_mod_gauss} shows a plot of real and imaginary part of both spin components of both spin Wannier states, $\tilde{\phi}_{1\uparrow}(\varphi)$ and $\tilde{\phi}_{1\downarrow}(\varphi)$, on site $n=1$, for potential strength $V_0 = 10\epsilon$ and Rashba coupling  $\alpha = 1.5$, calculated numerically on a grid with $N_{grid} = 240$ sites. As we can see, for both functions one spin component is dominant and the other one is negligible, which is what we expect from spin basis. This is the case even though the width of the functions is quite large compared to the inter-site spacing, which indicates that the choice of coefficients equation \eqref{eq:app_coefnss} gives good results even when the assumptions taken in their derivation are not fulfilled.

Spin Wannier states on figure \ref{fig:wann_mod_gauss} are also compared with the bound state $\tilde{\eta}_{ns}(\varphi)$ in a single Gaussian potential well equation \eqref{eq:Wgauss} of the same depth and width, which is relevant for the transition of electron between sites, further discussed in \ref{app:Zenner}.

\begin{figure}[ht]
\centerline{\includegraphics[scale=1]{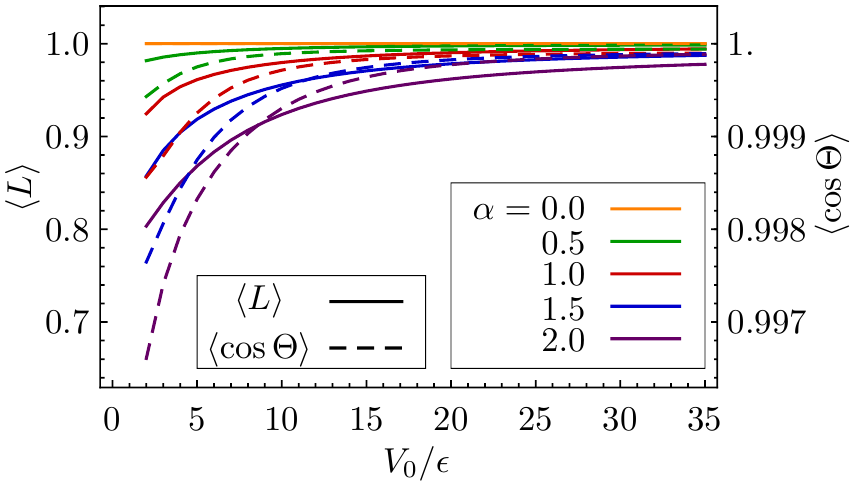}}
\caption[Wannier basis spin]{The length of vector of expectation values of spin and its angle from the $z$ axis for spin Wannier basis states $\tilde{\phi}_{n s}$ as a function of potential strength $V_0$ for different values of Rashba coupling.
}
\label{fig:spin_mod_wann}
\end{figure}

To verify that the spin properties of spin Wannier basis $\left| \tilde{\phi}_{ns}\right\rangle$ correspond to criterion equation \eqref{eq:sns_sWann}, we numerically calculate the expectation values of all three spin components
\begin{equation}
\left \langle \vec{s} \right\rangle = \left\langle \tilde{\phi}_{ns} \right| \vec{s} \left| \tilde{\phi}_{ns}\right\rangle \equiv \left( \left \langle s_x \right\rangle, \left \langle s_y \right\rangle, \left \langle s_z \right\rangle \right).
\end{equation}
To compare the spin properties of spin Wannier basis with that of pure spin state, we calculate the normalized length of the vector $\left \langle \vec{s} \right\rangle$ and the cosine of the angle that vector $\left \langle \vec{s} \right\rangle$ spans with the $z$-axis:
\begin{equation}
\left\langle L \right\rangle =  2 \left| \left\langle \vec{s} \right \rangle \right| /\hbar, \quad \left\langle \cos \Theta \right\rangle = \frac{\left \langle s_z \right\rangle}{\left| \left\langle \vec{s} \right \rangle \right|}.
\end{equation}
For pure spin state, both values are unity. Numerical calculated values of both quantities for a state $\left| \tilde{\phi}_{1 \uparrow}\right\rangle$ at same $N$, $\sigma$ and $N_{grid}$ as used for figure \ref{fig:wann_mod_gauss} are plotted in figure \ref{fig:spin_mod_wann} as a function of potential strength $V_0$ for various values of $\alpha$.

As seen in figure \ref{fig:spin_mod_wann}, in the absence of SO coupling, numerically calculated $\left\langle L \right\rangle$ and $\left\langle \cos \Theta \right\rangle$ are both $1$, , which indicates that in this limit, spin Wannier basis states $\left| \tilde{\phi}_{ns}\right\rangle$ are actually pure spin states. When the Rashba coupling is present, the parameters are no longer exactly one, but quickly approach this value when potential is increasing, indicating that spin Wannier basis, obtained with coefficients $c_{n s s'}$ equation \eqref{eq:app_coefnss} is indeed a very good approximation for pure spin states.

\section{Landau-Zenner transitions}
\label{app:Zenner}

Here we discuss the procedure of transferring the electron between two neighboring ring sites by changing the depth of local potential minimum.

As we see in figure \ref{fig:wann_mod_gauss} the spin Wannier basis functions $\left| \tilde{\phi}_{ns}\right\rangle$, calculated with coefficients equation \eqref{eq:app_coefnss}, are in fact very similar in shape to the bound states of the electron in the potential, consisting of only one potential well, labelled $\left| \tilde{\eta}_{ns}\right\rangle$.  We therefore assume for the rest of the discussion that the spin Wannier states and bound states are equivalent and that $\left| \tilde{\phi}_{ns}\right\rangle$ is also a stationary state of the potential with single potential minima at site $n$.

The Landau-Zenner transition between neighboring potential minima is realized in the following manner. We assume the initial potential on a ring to be a single potential minimum at site $n$,
\begin{equation}
V(\varphi,t=0) =  W(\varphi - n \varphi_a).
\end{equation}
with the electron initially in a superposition of spin Wannier basis states on the same site
\begin{equation}
\left| \psi_{init} \right\rangle = \sum_s c_s \left| \tilde{\phi}_{n s} \right \rangle.
\end{equation}
We then start to slowly decrease the depth of the potential at site $n$ and increase the depth at site $n+1$,
\begin{equation}
V(\varphi,t) = (1 - \beta t) W(\varphi - n \varphi_a) + \beta t W(\varphi - (n+1) \varphi_a).
\end{equation}
If voltage change rate $\beta$ is small $\hbar \beta \ll V_0$, this results in slow transition of electron from the superposition of spin Wannier states on site $n$ to the superposition of states on site $n+1$ \cite{Wittig2005},
\begin{equation}
\left| \psi(t) \right\rangle = \sum_s c_s(t) \left| \tilde{\phi}_{n s} \right \rangle + \sum_s d_s(t) \left| \tilde{\phi}_{n+1, s} \right \rangle.
\end{equation}
The probability of finding the electron on site $n$ or $n+1$ depends on magnitude of $c_s(t)$ and $d_s(t)$,
\begin{equation}
P_n(t) = \sum_s |c_s(t)|^2, \quad P_{n+1}(t) = \sum_s |d_s(t)|^2,
\end{equation}
and during slow transition, the value $P_n$ will change from $1$ to $0$ and $P_{n+1}$ from $0$ to $1$.

What is important for the spin transformation is the relation between coefficients of state in spin Wannier basis before ($c_n$) and after ($d_n$) electron transition. The lowest order term of time evolution operator $T(t) = \exp (-i \frac{H t}{\hbar})$, coupling the states $\left| \tilde{\phi}_{n s} \right\rangle$ and $\left| \tilde{\phi}_{n+1, s} \right\rangle$, is proportional to hopping matrix $\tilde{t}_{n s s'}^+$ equation \eqref{eq:hoppingt}. The state after the Landau-Zenner transition is also normalized, which leads us to the prediction that the coefficients of the final state in spin Wannier basis are related to initial coefficients as
\begin{equation}
\label{eq:d_s_app}
d_s = \sum_{s'} \frac{1}{t_0} \tilde{t}_{n s s'}^+ c_{s'}.
\end{equation}

\begin{figure}[ht]
\centerline{\includegraphics[scale=1]{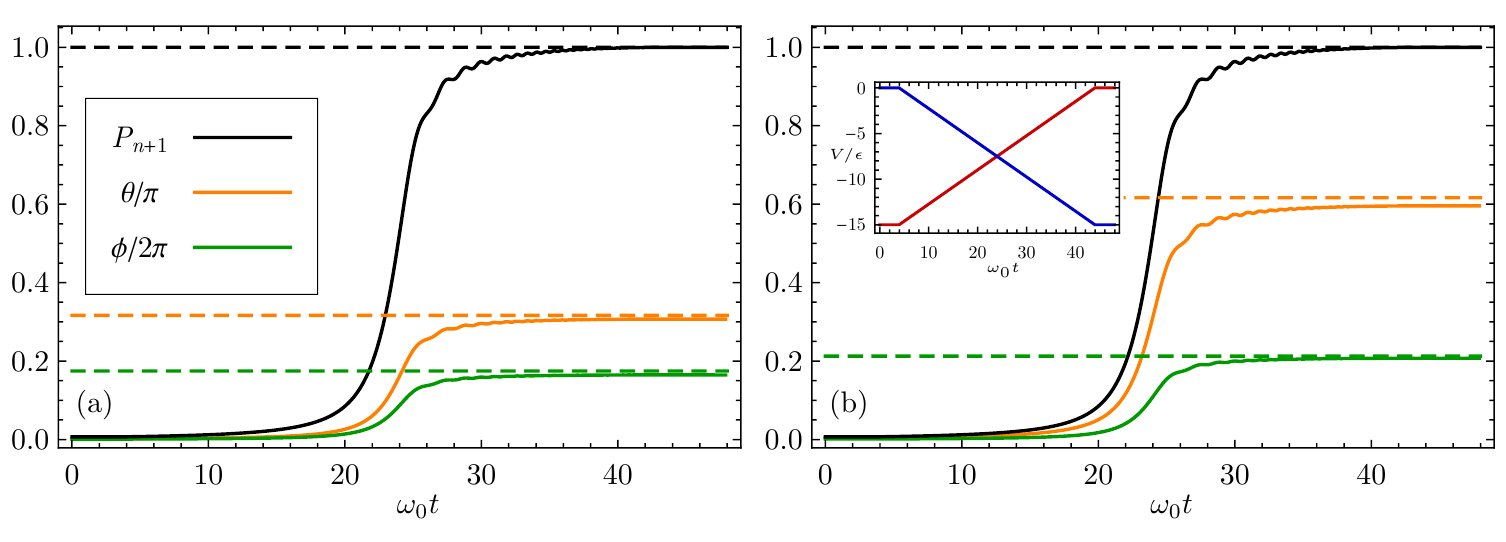}}
\caption[Zenner transition]{Numerically calculated time dependence of parameters $P_{n+1}(t)$, $\theta_d(t)$ and $\phi_d(t)$ during Landau-Zenner transition (solid lines), compared with analytically predicted results after transition (dashed line), for the transition between sites $n=3$ and $n+1 =4$. Panel a) shows the case of Rashba coupling $\alpha = 1$ and panel b) for $\alpha =2$. The transition was calculated in the potential of Gaussian shape equation \eqref{eq:Wgauss} with inter-site distance $\varphi_a = 2\pi/6$, with the initial depth of the potential well $V_0 = 15 \epsilon$ and width $\sigma = 0.1$ on a computational grid of $N_{grid}=90$ sites. The rate of potential change was set to $\beta = \frac{\omega_0}{40}$, where $\omega_0 = \epsilon/\hbar$ is a natural frequency of the system. Time dependency of local potential on sites $n$ and $n+1$ is shown on inset figure in panel (b) in red and blue, respectively.}
\label{fig:Zenner_trans}
\end{figure}

We verified this result by numerical calculation of the coefficients $d_\uparrow(t)$ and $d_\downarrow(t)$. The results are presented in figure \ref{fig:Zenner_trans} as the probability $P_{n+1}(t)$ of finding the electron on site $n+1$, and the direction of a vector of expectation values of Pauli matrices, calculated from coefficients $d_s$,
\begin{equation}
\left\langle \sigma_i \right\rangle_d(t) = \sum_{s s'} d_s^*(t) \sigma_{i s s'} d_{s'}(t),
\end{equation}
expressed by angle $\theta_d$ and $\phi_d$:
\begin{equation}
\eqalign{
\theta_d(t) = \arccos \left( \frac{ \left\langle \sigma_z \right\rangle_d(t)}{ \left| \left\langle \vec{\sigma} \right\rangle_d \right|(t)} \right), \quad
\phi_d(t) = \arctan \left( \frac{ \left\langle \sigma_y \right\rangle_d (t)}{ \left\langle \sigma_x \right\rangle_d (t)}  \right).
}
\end{equation}
The values $P_{n+1}(t)$ and $\theta_d(t)$ and $\phi_d(t)$ determine the coefficients $d_{ns}(t)$ up to a complex phase and therefore contain all physically relevant information. Since the spin Wannier states are basically equivalent to the pure spin states (see \ref{app:WannSpinBase}), the expectation values $\left\langle \sigma_i \right\rangle$ are closely related to the actual spin expectation values,
\begin{equation}
\left\langle s_i \right\rangle = \left\langle\psi \right| s_i \left| \psi \right\rangle \approx \frac{\hbar}{2} \left\langle \sigma_i \right\rangle_d.
\end{equation}
By plotting the values $P_{n+1}(t)$ and $\theta_d(t)$ and $\phi_d(t)$ we therefore extract all physically relevant information about electron's position and its spin orientation. 

The time dependence of relevant quantities is plotted as solid lines in figure \ref{fig:Zenner_trans}. The dashed lines are the values, calculated from the coefficients $d_s$, predicted in equation \eqref{eq:d_s_app}, which result in expectation values of the Pauli vector 
\begin{equation}
\left\langle \vec{\sigma} \right\rangle = \left( 
\begin{array}{c}
2 s_{\alpha } s_{\phi } \left(s_{\varphi_a} c_{\phi }-c_{\varphi_a} c_{\alpha } s_{\phi }\right)\\
2 s_{\alpha } s_{\phi } \left(s_{\varphi_a} c_{\alpha } s_{\phi }+c_{\varphi_a} c_{\phi
   }\right)\\
   c_{\alpha }^2+s_{\alpha }^2 \left(c_{\phi }^2 - s_{\phi } ^2\right) 
\end{array}
 \right),
\end{equation}
with $s_\alpha$, $c_\alpha$, $s_\phi$ and $c_\phi$ defined in equation \eqref{eq:sc_def} and
\begin{equation}
s_{\varphi_a} = \sin \varphi_a, \quad c_{\varphi_a} = \cos \varphi_a.
\end{equation}

From figure \ref{fig:Zenner_trans} it is evident that the numerical results agree very well with the theoretical prediction, from which we conclude that the equation for coefficients of state in spin Wannier basis after the Landau-Zenner transition equation \eqref{eq:d_s_app} is indeed a good approximation for the analysis of spin transformations.


\section*{References}

\begin{thebibliography}{10}
\bibitem{Wolf2001}
Wolf S~A, Awschalom D~D, Buhrman R~A, Daughton J~M, von Moln\'{a}r S, Roukes
  M~L, Chtchelkanova A~Y and Treger D~M 2001 {\em Science\/} {\bf 294} 1488

\bibitem{Zutic2004}
\v{Z}uti\'{c} I and {Das Sarma} S 2004 {\em Rev. Mod. Phys.\/} {\bf 76}
  323

\bibitem{Rashba2007}
Rashba E~I 2007 {\em {Future Trends in Microelectronics}\/} (Hoboken, NJ, USA:
  John Wiley \& Sons, Inc.)
  
\bibitem{Awschalom2013}
Awschalom D~D, Bassett L~C, Dzurak A~S, Hu E~L and Petta J~R 2013 {\em
  Science\/} {\bf 339} 1174

\bibitem{Winkler2003}
Winkler R 2003 {\em {Spin--Orbit Coupling Effects in Two-Dimensional Electron
  and Hole Systems}\/} ({\em Springer Tracts in Modern Physics\/} vol 191)
  (Berlin, Heidelberg: Springer Berlin Heidelberg)

\bibitem{Engel2007}
Engel H~A, Rashba E~I and Halperin B~I 2007 {\em {Handbook of Magnetism and
  Advanced Magnetic Materials}\/} (Chichester, UK: John Wiley \& Sons, Ltd)
  
\bibitem{Rashba1960}
Rashba E~I 1960 {\em Sov. Phys. Solid State\/} {\bf 2} 1109

\bibitem{Nitta1997}
Nitta J, Akazaki T, Takayanagi H and Enoki T 1997 {\em Phys. Rev. Lett.\/} {\bf
  78} 1335
  
\bibitem{Schapers1998}
Schapers T, Engels G, Lange J, Klocke T, Hollfelder M and Luth H 1998 {\em
  J Appl. Phys.\/} {\bf 83} 4324

\bibitem{Datta1990}
Datta S and Das B 1990 {\em Appl. Phys. Lett.\/} {\bf 56} 665

\bibitem{Nitta1999}
Nitta J, Meijer F~E and Takayanagi H 1999 {\em Appl. Phys. Lett.\/} {\bf 75}
  695

\bibitem{Schliemann2003}
Schliemann J, Egues J~C and Loss D 2003 {\em Phys. Rev. Lett.\/} {\bf 90}
  146801 

\bibitem{Wunderlich2010}
Wunderlich J, Park B~G, Irvine A~C, Z\^{a}rbo L~P, Rozkotov\'{a} E, Nemec P,
  Nov\'{a}k V, Sinova J and Jungwirth T 2010 {\em Science\/} {\bf 330} 1801
\bibitem{Stajic2013}
Stajic J 2013 {\em Science\/} {\bf 339} 1163

\bibitem{Ramsak2018}
Ram{\v{s}}ak A, {\v{C}}ade{\v{z}} T, Kregar A and Ul{\v{c}}akar L 2018 {\em The
  European Physical Journal Special Topics\/} {\bf 227} 353

\bibitem{Cadez2013}
\v{C}ade\v{z} T, Jefferson J~H and Ram\v{s}ak A 2013 {\em New J. Phys.\/}
  {\bf 15} 013029
  
\bibitem{Cadez2014}
\v{C}ade\v{z} T, Jefferson J~H and Ram\v{s}ak A 2014 {\em Phys. Rev. Lett.\/}
  {\bf 112} 150402

\bibitem{Ulcakar2017}
Ul\v{c}akar L and Ram\v{s}ak A 2017 {\em New J. Phys.\/}
  {\bf 19} 093015
  
\bibitem{Donvil2020}
Donvil B, Ul\v{c}akar L, Rejec T, and Ram\v{s}ak A, arXiv:2002.05548.

\bibitem{San-Jose2008}
San-Jose P, Scharfenberger B, Sch\"{o}n G, Shnirman A and Zarand G 2008 {\em
  Phys. Rev. B\/} {\bf 77} 045305

\bibitem{Golovach2010}
Golovach V~N, Borhani M and Loss D 2010 {\em Phys. Rev. A\/} {\bf 81} 022315
 
\bibitem{Meijer2002}
Meijer F, Morpurgo A and Klapwijk T 2002 {\em Phys. Rev. B\/} {\bf 66} 033107

\bibitem{Kregar2016}
Kregar A, Jefferson J~H and Ram\v{s}ak A 2016 {\em Phys. Rev. B\/} {\bf
  93} 1

\bibitem{Kittel2004}
Kittel C 2004 {\em Introduction to Solid State Physics\/} 8th ed (Wiley)

\bibitem{Kregar2016b}
Kregar A and Ram{\v{s}}ak A 2016 {\em Int. J. Mod. Phys. B\/} {\bf 30} 1642016

\bibitem{Thalineau2012}
Thalineau R, Hermelin S, Wieck A~D, Bauerle C, Saminadayar L and Meunier T
  2012 {\em Appl. Phys. Lett\/} {\bf 101} 103102

\bibitem{Grundler2000}
Grundler D 2000 {\em Phys. Rev. Lett\/} {\bf 84} 6074

\bibitem{Premasiri2019}
Premasiri K and Gao X~P~A 2019 {\em J. Phys. Condens. Matter\/}
  {\bf 31} 19300

\bibitem{Liang2012}
Liang D and Gao X~P~A 2012 {\em Nano Letters\/} {\bf 12} 3263

\bibitem{Bayat2013}
Bayat A, Creffield C~E, Jefferson J~H, Pepper M and Bose S 2015 {\em Semicond.
  Sci. Tech.\/} {\bf 30} 105025

\bibitem{Wittig2005}
Wittig C 2005 {\em J. Phys. Chem. B\/} {\bf 109} 8428


\end{thebibliography}

\end{document}